\documentclass[aps,pra,reprint,showpacs,groupedaddress]{revtex4-1}
\usepackage[final]{graphicx}

\begin{document}


\title{Quantum Optimal Control Theory in the Linear Response Formalism}


\author{Alberto Castro}
\email[]{acastro@bifi.es}
\affiliation{Institute for Biocomputation and Physics of Complex Systems (BIFI) and 
Zaragoza Center for Advanced Modelling (ZCAM), University of Zaragoza, Spain}

\author{I. V. Tokatly}
\email[]{ilya_tokatly@ehu.es}
\affiliation{Nano-Bio Spectroscopy group and ETSF Scientific Development Centre, 
  Departamento de F\'isica de Materiales, Universidad del Pa\'is Vasco UPV/EHU, E-20018 San Sebasti\'an, Spain}
\affiliation{IKERBASQUE, Basque Foundation for Science, E-48011 Bilbao, Spain}



\date{\today}

\begin{abstract}
  Quantum optimal control theory (QOCT) aims at finding an external
  field that drives a quantum system in such a way that optimally
  achieves some predefined target. In practice this normally means
  optimizing the value of some observable, a so called merit
  function. In consequence, a key part of the theory is a set of
  equations, which provides the gradient of the merit function with
  respect to parameters that control the shape of the driving
  field. We show that these equations can be straightforwardly derived
  using the standard linear response theory, only requiring a minor
  generalization -- the unperturbed Hamiltonian is allowed to be
  time-dependent. As a result, the aforementioned gradients are
  identified with certain response functions. This identification
  leads to a natural reformulation of QOCT in term of the Keldysh
  contour formalism of the quantum many-body theory. In particular,
  the gradients of the merit function can be calculated using the
  diagrammatic technique for non-equilibrium Green's functions, which
  should be helpful in the application of QOCT to computationally
  difficult many-electron problems.
\end{abstract}



\maketitle

\section{Introduction}

Quantum Optimal Control Theory
(QOCT)~\cite{Brif2010,doi:10.1088/0953-4075/40/18/R01} is concerned
with finding a time-dependent external field that drives a given
quantum system to optimally achieve some predefined target, that
depends on the manner in which the system evolves~\footnote{Perhaps
  more generally, the optimization can also be done with respect to
  \emph{internal} parameters defining the system itself, and not
  necessarily defining an external field. However, most applications
  are for this latter case}. For example, a target can be the
population of some excited state at the final time of the propagation
-- but many other options are possible. The theory can be regarded as
a branch of the \emph{classical} control theories developed mostly in
the fields of mathematics and
engineering~\cite{luenberger1969,luenberger1979}. The quantum
discipline was born in the late
80s~\cite{shi19886870,peirce19884950,kosloff1989201}, as the most
complete theoretical framework capable of addressing the nascent
experimental field of quantum control (or \emph{coherent
  control})~\cite{shapiro-2003}. The range of applications of QOCT is
growing very fast, thanks to the progress in the ultrafast laser pulse
generation and pulse shaping techniques~\cite{weiner2000}, as well as
to the development of adaptive feedback control
schemes~\cite{judson1995,bardeen1997}. Typical examples of
applications are the control of the population of excited states in
molecules~\cite{bardeen1997}, optimization of high-harmonic
generation~\cite{bartels2000}, optimization of selective
photo-dissociation of molecules~\cite{Assion30101998}, optimization of
multi-photon ionization of atoms~\cite{PhysRevLett.92.208301},
enhancement of electron transfer in dye-sensitized solar
cells~\cite{PhysRevLett.97.208301}, etc.

At the formal level, the central problem of QOCT is to maximize an
expectation value of some operator, usually known as a metrit (or
target) function, whose input is the external field that needs to be
optimally shaped. The field is normally parameterized either by a
decrete set of real-valued ``control'' parameters, or, in a more
general setting, by continuous functions of time. In the latter case,
one usually speaks of target {\rm functionals}.
In most cases, the optimization algorithm will require both the
computation of the merit function and of its gradient with respect to
control parameters. Therefore an expression and computational strategy
for this gradient constitutes one of the most important parts of QOCT.

The usual derivation of expressions for the gradient of the merit
function proceeds via the definition of a Lagrangian functional, and
of a ``Lagrange multiplier'' wave function (see, for example,
Refs.~\cite{peirce19884950,tersigni-1990}). It leads to an expression
for the gradient that involves the forward propagation of the system
wave function, and the backwards propagation of the new Lagrange
multiplier wave function. At this point it is worth noting that the
presence of forward and backwards time progations is a general feature
of the quantum kinetic theory which can be conveniently formulated as
a propagation along the Keldysh-Schwinger closed-time contour
\cite{Keldysh1965,Schwinger1961}. Therefore it is natural to expect
that there is a connection between QOCT and the Keldysh contour
formulation of the quantum dynamics. In the present work we make this
conection explicit by re-examining the derivation of the expression of
the gradient (or functional derivative) of the target functional.

Our main simple observation is that the differentiation of a target
observable with respect to a control parameter is identical to
computing a change of that observable induced by a corresponding
perturbation in the Hamiltonian. Thus, the problem of calculating the
gradient of the merit function reduces to a generalized form of linear
response theory (LRT), in which the unperturbed Hamiltonian is no
longer static but depends on time. The formalism of LRT can then be
directly applied, and we straightforwardly recover the very same
expressions that one reaches in the ``traditional'' way. However,
these expressions can then be regarded as response functions
represented by certain retarded correlation functions.  We emphasize
that this re-derivation is not a mere academic exercise, since the new
interpretation of the gradient as a response function suggests
immediately the use of the known approximations to this object. In
particular by relating the retarded response function to a
contour-ordered correlation function we can apply well developed
methods and approximations of the non-equilibrium many-body
perturbation theory to QOCT for many-electron systems
\cite{Keldysh1965,KadBay,Danielewicz1984}.

The latter is a specially important aspect, since the treatment of
many-electron systems in notoriously difficult; yet the direct control
of electrons is an area of growing interest, due to the advances in
laser pulses of strong intensity and ultra-short durations, in the
atto-second range -- the scale of the electronic movements. In order
to theoretically study a direct control of electronic motion, it is
necessary to have a predictive (\emph{ab initio}) yet computational
tractable scheme, in combination with QOCT. Some possibilities have
been recently put forward, such as (multi-configuration)
time-dependent Hartree Fock~\cite{doi:797291} and time-dependent
density-functional theory~\cite{arXiv:1009.2241}. Here we propose a
new possibility, based on non-equilibrium many-body Green's functions
theory.

The structure of the paper is the following. In
Sec.~\ref{section:rederivation}, we derive the gradient QOCT equations
in the formalism of LRT. To make this paper self-contained, the
slightly generalized basic LRT results needed for this purpose are
presented in Appendix~\ref{appendix-lrt}. Sec.~\ref{śection:negft}
elaborates on the equations derived in Sec.~\ref{section:rederivation}
by proposing a QOCT scheme for many-body systems, based on the Keldysh
contour formalism and on standard approximations in non-equilibrium
many-body Green's functions theory.

\section{The basic QOCT equations in the linear response theory language}
\label{section:rederivation}

Let us consider a quantum system described by its density matrix
$\hat{\rho}(t)$ and governed, in the time interval $[t_0,t_f]$, by a
von Neumann equation in the form:
\begin{eqnarray}
\frac{\partial}{\partial t}\hat{\rho}(t) & = & -i \left[ \hat{H}[u](t), \hat{\rho}(t)\right]\,,
\\
\hat{\rho}(t_0) & = & \hat{\rho}_0\,,
\end{eqnarray}
where the Hamiltonian is given by~\footnote{Eq.~(\ref{eq:hamiltonian})
  prescribes a particular form of the Hamiltonian, namely a linear
  dependence of it with the control field $\epsilon[u]$. This is
  perhaps the most common case, and may describe, for example, a laser
  pulse interacting with an atom or molecule. However, the results
  that follow do not rely on this particular choice, and could be
  derived with a generic dependence $\protect\hat{H}[u]$.}:
\begin{equation}
\label{eq:hamiltonian}
\hat{H}[u](t) = \hat{\mathcal{H}} + \epsilon[u](t)\hat{V}\,.
\end{equation}
The Hamiltonian piece $\hat{\mathcal{H}}$ is static, and
$\epsilon[u](t)$ is a time-dependent function whose precise form is
determined by a set of parameters that we will denote, collectively,
$u$. The operator $\hat{V}$ represents the coupling of the system with
an external field, e.g. if we think of an atom or molecule irradiated
by a laser pulse, the dipole operator. Evidently, a particular choice
of the \emph{control} $u$ leads to a system evolution, $u \rightarrow
\hat{\rho}[u](t)$.

We wish to find the values of $u$ that maximize the value of the
expectation value of some observable $\hat{A}$ at the end of the
propagation. In other words, we want to find the maximum of the function:
\begin{equation}
G[u] = {\rm Tr} \{ \hat{\rho}[u](t_f)\hat{A}\}\,.
\end{equation}
In order to find the maximum, the best way is to be
able to compute the gradient of $G$. The problem that we face,
therefore, is that of finding a suitable expression for this gradient.

Assuming that there is only one parameter $u$ (the generalization to more than one is trivial):
\begin{equation}
\frac{\partial G}{\partial u}[u] = \lim_{\Delta u \to 0}\Delta u^{-1}(G[u+\Delta u] - G[u])\,.
\end{equation}
Note that $\hat{\rho}[u]$ corresponds to the propagation of the system
with the Hamiltonian given in Eq.~(\ref{eq:hamiltonian}), whereas
$\hat{\rho}[u+\Delta u]$ corresponds to the propagation of the system
with the Hamiltonian
\begin{equation}
\hat{H}[u+\Delta u](t) = 
\hat{H}[u](t) + \Delta u\frac{\partial \epsilon}{\partial u}[u]\hat{V}\,,
\end{equation}
to first order in $\Delta u$. Now we can use directly the LRT result
introduced in appendix \ref{appendix-lrt}, by making the identifications:
\begin{equation}
\hat{H}_0(t) =  \hat{H}[u](t),\quad f(t) =  \Delta u\frac{\partial \epsilon}{\partial u}[u](t)\,.
\end{equation}
Therefore, we just need to apply
Eqs.~(\ref{eq:lrdef}) and (\ref{eq:deltaa}) to arrive at:
\begin{equation}
\label{eq:gradient}
\frac{\partial G}{\partial u}[u] = \int_{t_0}^{\infty}\!\!\!\!{\rm d}\tau\; \frac{\partial \epsilon}{\partial u}[u](\tau)
\chi_{\hat{A},\hat{V}}(t_f,\tau)\,.
\end{equation}
where
\begin{equation}
\label{eq:chi-general}
\chi_{\hat{A},\hat{V}}(t_f,\tau) = -i \theta(t_f-\tau)
{\rm Tr}\{ \hat{\rho}(t_0) \left[ \hat{A}_H(t_f),\hat{V}_H(\tau)\right]\}
\end{equation}
is the response function for the $(\hat{A},\hat{V})$ operators. Inside the commutator, these operators
appear in the Heisenberg representation, defined by:
\begin{equation}
\hat{O}_H(t) = \hat{U}^\dagger(t,t_0)\hat{O}\hat{U}(t,t_0)
\end{equation}
for any observable $\hat{O}$, and where $\hat{U}(t,t_0)$ is the
propagator corresponding to the $\hat{H}[u](t)$
Hamiltonian. Eq.~(\ref{eq:gradient}) clearly manifests how the
gradient is nothing else than a response function -- albeit a
generalized one. It corresponds to the response of a system driven by
a time-dependent Hamiltonian, to a modification of this
Hamiltonian. It remains now to see how this result is equivalent to
the expressions obtained in a different manner with the usual QOCT
technique. For that purpose, we define an operator:
\begin{eqnarray}
\hat{A}[u](\tau) & = & \hat{U}(\tau,t_f)\hat{A}\hat{U}(t_f,\tau)\,,
\\
\hat{A}[u](t_f) & = & \hat{A}\,,
\end{eqnarray}
which can also be written as the solution to the differential equation:
\begin{eqnarray}
\label{eq:a}
\frac{\partial}{\partial t}\hat{A}[u](t) & = & -i\left[ \hat{H}[u](t), \hat{A}[u](t)\right]\,,
\\
\label{eq:aT}
\hat{A}[u](t_f) & = & \hat{A}\,.
\end{eqnarray}
Using this new auxiliary object, and after a little manipulation of
Eq.~(\ref{eq:gradient}) one arrives at:
\begin{equation}
\label{eq:mainqocteq}
\frac{\partial G}{\partial u}[u] = - i
\int_{t_0}^{t_f}\!\!\!\!{\rm d}\tau\; \frac{\partial \epsilon}{\partial u}[u](\tau)
{\rm Tr}\{ \hat{\rho}[u](\tau) \left[ \hat{A}[u](\tau),\hat{V}\right]\}\,.
\end{equation}
Equations~(\ref{eq:a}), (\ref{eq:aT}) and (\ref{eq:mainqocteq}),
together with the original propagation equation for
$\hat{\rho}[u](t)$, are the ``QOCT equations'', usually derived in a
different way (through the definition of a Lagrangian
function). Algorithmically, the computation of the gradient is
performed with two consecutive propagations, one forwards for the
original system equations, and one backwards in order to obtain
$\hat{A}[u](t)$. These propagations provide the necessary ingredients
to compute Eq.~(\ref{eq:mainqocteq}). In the next section we will make
a link of these forward and backwards propagations to the formulation
of the quantum dynamics via the Keldysh contour formalism.

It is also easy to see that all variations and generalizations of the
QOCT equations naturally follow from our linear response approach.

\subsection{Pure states}

For the case of a pure state dynamics the density matrix takes the form $\hat{\rho}[u](t) = \vert\Psi[u](t)\rangle\langle\Psi[u](t)\vert$, where the wave function $\vert\Psi[u](t)\rangle$ evolves from a given initial state $\quad \vert\Psi[u](t_0)\rangle = \vert\Psi_0\rangle$ according to the Schr\"odinger equation
\begin{equation}
 \label{eq:SE}
\frac{\partial}{\partial t}\vert\Psi[u](t)\rangle = -i \hat{H}[u](t)\vert\Psi[u](t)\rangle.
\end{equation}
The gradient of the merit function is given by the general Eq.~(\ref{eq:gradient}). The only difference is that now the initial density matrix entering the response function describes a pure state: $\hat{\rho}_0 = \vert\Psi_0\rangle\langle\Psi_0\vert$. Hence Eq.~(\ref{eq:chi-general}) reduces to the form
\begin{equation}
\label{eq:chi-pure}
\chi_{\hat{A},\hat{V}}(t,\tau) = -i \theta(t-\tau)
\langle\Psi_0\vert\left[ \hat{A}_H(t),\hat{V}_H(\tau)\right]\vert\Psi_0\rangle.
\end{equation}
Inserting this equation into Eq.~(\ref{eq:gradient}), writing the commutator explicitly, and inspecting the terms we find that the gradient can be written as follows
\begin{equation}
\label{eq:qoct-gradient-1}
\frac{\partial G}{\partial u}[u] = 2{\rm Im}
\int_{t_0}^{t_f}\!\!\!\!{\rm d}\tau\; \frac{\partial \epsilon}{\partial u}[u](\tau)
\langle \chi[u](\tau)\vert\hat{V}\vert\Psi[u](\tau)\rangle\,.
\end{equation}
where $\vert\chi[u](t)\rangle$ is defined by the expression
\begin{equation}
 \vert\chi[u](t)\rangle = \hat{U}(t,t_f)\hat{A}\vert\Psi[u](t_f)\rangle.
\end{equation}
Alternatively this function can be viewed as a solution to the following backwards propagation problem
\begin{eqnarray}
\label{eq:chi}
\frac{\partial}{\partial t}\vert\chi(t)\rangle & = & -i\hat{H}[u](t)\vert\chi(t)\rangle\,,
\\
\label{eq:chiT}
\vert\chi(t_f)\rangle & = & \hat{A}\vert \Psi[u](t_f)\rangle\,,
\end{eqnarray}
which coincides with the standard QOCT equations for pure
states. Within the usual formalism the state $\vert\chi[u](t)\rangle$
appears as a ``Lagrange multiplier'' wave function.


\subsection{Continuous parameters}

The case in which the control function $\epsilon(t)$ is not
parameterized, but one does the search in the whole space of
continuous functions, can also be treated essentially in the same
manner. In this case, instead of a gradient we will obtain a
functional derivative; in fact, this derivative is nothing else than
the response function, i.e. Eq.~(\ref{eq:gradient}) is simply:
\begin{equation}
\frac{\delta G}{\delta \epsilon(t)} = \chi_{\hat{A},\hat{V}}(t_f,t)\,.
\end{equation}
This can be rewritten, for the pure state case, as:
\begin{equation}
\frac{\delta G}{\partial \epsilon(t)} = 2{\rm Im}
\langle \chi[\epsilon](t)\vert\hat{V}\vert\Psi[\epsilon](t)\rangle\,,
\end{equation}
where $\chi[\epsilon](t)$ is the solution to Eqs.~(\ref{eq:chi}) and (\ref{eq:chiT}).

\subsection{General target functionals}

In some cases, the function to optimize is not a simple
expectation value of an operator $\hat{A}$, but perhaps a more
general expression in the form:
\begin{equation}
G[u] = F[\hat{\rho}[u],u]\,,
\end{equation}
where $F$ is a functional of the evolution of the system (and also
perhaps explicitly of the control parameters, hence the second
argument). Normally, this is split as:
\begin{equation}
G[u] = J_1[\hat{\rho}[u]] + J_2[u]\,,
\end{equation}
i.e. the first term is the \emph{real} objective, depending on the
evolution of the system, whereas the second term is added in order
to penalize undesired features of the control function, such as for
example too high frequencies or intensities. In any case, any
physically meaningful definition for $J_1$ will be that in which it
is a function of expectation values of observables. In this case the
derivation outlined here is directly applicable, by a simple use of
the chain rule.

\subsection{Time-dependent targets}

A more interesting generalization is that in which the function
to optimize depends on the expectation value of the operator at all
times during the propagation, and not only at the final time $t_f$:
Once again, this case can also be put in response-function language
in a rather straightforward manner. Let us consider for example the
pure-state case:
\begin{equation}
G[u] = \int_{t_0}^{t_f}\!\!\!\!{\rm d}t\; g(t) \langle \Psi[u](t)\vert\hat{A}\vert\Psi[u](t)\rangle\,,
\end{equation}
where $g(t)$ is some weight function. The application of the LRT equations leads now to:
\begin{equation}
\label{eq:gradient-td}
\frac{\partial G}{\partial u}[u] = \int_{t_0}^{t_f}\!\!\!\! {\rm d}t\;
\int_{t_0}^{\infty}\!\!\!\!{\rm d}\tau\; g(t)\frac{\partial \epsilon}{\partial u}[u](\tau)
\chi_{\hat{A},\hat{V}}(t,\tau)\,.
\end{equation}
Here the response function $\chi_{\hat{A},\hat{V}}(t,\tau)$ is given by Eq.~(\ref{eq:chi-pure}). Following the same route as in derivation of Eq.~(\ref{eq:qoct-gradient-1}) in Sec.~IIIA we rewrite Eq.~(\ref{eq:gradient-td}) as:
\begin{equation}
\frac{\partial G}{\partial u}[u] = 2{\rm Im} \int_0^{t_f}\!\!\!\!{\rm d}\tau\;
\frac{\partial \epsilon}{\partial u}[u](\tau) \langle \chi[u](\tau)\vert\hat{V}\vert\Psi[u](\tau)\rangle\,,
\end{equation}
where $\chi[u](\tau)$ is defined by the following integral
\begin{equation}
\vert\chi[u](\tau)\rangle = \int_{\tau}^{t_f}\!\!\!\!{\rm d}t\;g(t)\hat{U}(t,\tau)\hat{A}\vert\Psi[u](t)\rangle\,,
\end{equation}
which can be put in the equivalent differential form:
\begin{eqnarray}
\frac{\partial}{\partial \tau}\vert\chi[u](\tau)\rangle & = & - i\hat{H}[u](\tau)\vert\chi[u](\tau)\rangle - g(\tau)\hat{A}\vert\Psi[u](\tau)\rangle\,.
\\
\vert\chi[u](t_f)\rangle & = & 0\,.
\end{eqnarray}
These are once again the backwards QOCT equations, in the case of ``time-dependent targets''.

\section{QOCT in terms of the Keldysh contour formalism}
\label{śection:negft}

\begin{figure}
\setlength{\unitlength}{0.7\columnwidth}
\centerline{
\begin{picture}(1.0,0.2)
\put(0.0,0.1){\vector(1,0){1}}
\thicklines
\put(0.0,0.12){\vector(1,0){0.3}}
\put(0.3,0.12){\line(1,0){0.4}}
\put(0.7,0.10){\oval(0.06,0.04)[r]}
\put(0.7,0.08){\vector(-1,0){0.3}}
\put(0.4,0.08){\line(-1,0){0.4}}
\put(0.0,0.01){$t_0$}
\put(0.95,0.01){$t$}
\put(0.7,0.01){$t_f$}
\put(0.0,0.12){\circle*{0.02}}
\end{picture}
}
\caption{
\label{fig1}
Keldysh contour.
}
\end{figure}

The new point of view on QOCT proposed in the previous section
naturally suggests new approximation strategies for control problems
in interacting many-electron systems. As we will now show the QOCT
equations can be expressed in terms of correlations functions defined
on a Keldysh \cite{Keldysh1965} closed time contour. This allows for
an immediate application of the powerful machinery of non-equilibrium
Green's functions theory to the coherent control problem.

Let us reconsider the key equation for the gradient of the merit function, Eq.~(\ref{eq:gradient}), and write it explicitly as follows
\begin{eqnarray}\nonumber
\frac{\partial G}{\partial u}[u] &=& - i\int_{t_0}^{t_f}d\tau\; \frac{\partial \epsilon}{\partial u}[u](\tau)
{\rm Tr}\{ \hat{\rho}(t_0) \hat{A}_H(t_f)\hat{V}_H(\tau)\} \\
&-& i\int_{t_f}^{t_0}d\tau\; \frac{\partial \epsilon}{\partial u}[u](\tau)
{\rm Tr}\{ \hat{\rho}(t_0) \hat{V}_H(\tau)\hat{A}_H(t_f)\}.
\label{eq:gradient2}
\end{eqnarray}
The two integrals in this equation can be composed into a single
integral over the Keldysh contour $C$ depicted in
Fig.~(\ref{fig1}). This contour starts at $t_0$, goes froward in time
to $t_f$, and then comes back to the origin. Therefore by using the
standard definition of a contour-ordered correlation function
\begin{equation}
\label{eq:chi-contour}
\chi^{C}_{\hat{A},\hat{V}}(\tau,\tau') = - i {\rm Tr}\{ \hat{\rho}(t_0) {\rm T}_C \left[ \hat{A}_H(\tau)\hat{V}_H(\tau') \right] \}\,
\end{equation}
where ${\rm T}_C$ is the chronological ordering operator on the contour $C$, we can cast Eq.~(\ref{eq:gradient2}) into the following compact form
\begin{equation}
\label{eq:gradient-contour}
\frac{\partial G}{\partial u}[u] = \int_C d\tau \; \frac{\partial \epsilon}{\partial u}[u](\tau)
\chi^{C}_{\hat{A},\hat{V}}(t_f,\tau)\,,
\end{equation}

The main advantage of the representation (\ref{eq:gradient-contour})
is that for interacting many-body systems the contour-ordered
correlation functions can be calculated using the standard
diagrammatic technique for non-equilibrium Keldysh Green's functions
(see, e.~g,,
Refs.~\onlinecite{Keldysh1965,KadBay,Danielewicz1984,Danielewicz1990,DahStaLee2006,stan-2009,FriVerAlm2009}). In
other words, by employing the well developed machinery/approximations
of the non-equilibrium Green's functions theory (NEGFT) we can express
the gradients of the merit function as a functional of the contour
ordered one-particle Green's functions.

To illustrate above statements we consider the simplest situation when
both the control field $\hat{V}$ and the observable of interest
$\hat{A}$ are represented by one-particle operators. In this case the
correlation function $\chi^{C}_{\hat{A},\hat{V}}(t_f,\tau)$ entering
Eq.~(\ref{eq:gradient-contour}) is given by:
\begin{equation}
\setlength{\unitlength}{1pt}
\begin{picture}(200,40)
\put(0,17){$\chi_{\hat{A},\hat{V}}^C(t_f,\tau) =$}
\put(60,0){
\includegraphics[clip]{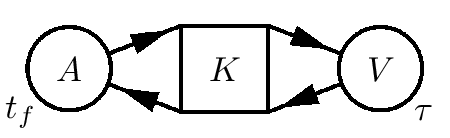}
}
\end{picture}
\end{equation}
where $K$ is the exact two-particle Green's
function. Now we can take our favorite many-body approximation, such
as Hartree-Fock, second-Born, $T$-matrix, random phase approximation
(RPA), etc., to get an explicit and practically feasible
expression. For example, at the RPA/GW level the correlation function
reduces to the two following terms:
\begin{equation}
\setlength{\unitlength}{1pt}
\begin{picture}(200,120)
\put(0,87){$\chi_{\hat{A},\hat{V}}^C(t_f,\tau) =$}
\put(60,60){
\includegraphics[clip]{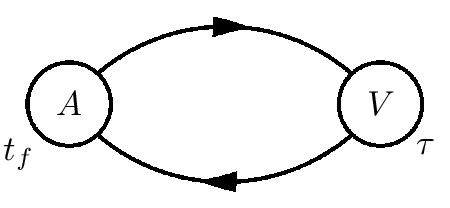}
}
\put(30,27){+}
\put(40,0){
\includegraphics[clip]{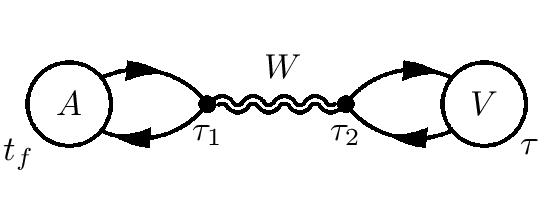}
}
\end{picture}
\end{equation}
Analytically, this diagram translates to:
\begin{eqnarray}
 \nonumber
&&\chi^{C}_{\hat{A},\hat{V}}(t_f,\tau) = {\rm tr}\{\hat{A}G(t_f,\tau)\hat{V}G(\tau,t_f)\}\\
\nonumber
&& + \int d\tau_1 d\tau_2\int d{\bf r_1}d{\bf r_2}{\rm tr}\{\hat{A}G(t_f,\tau)\hat{n}({\bf r_1})G(\tau,t_f)\}\\
&&\times W({\bf r_1},\tau_1;{\bf r_2},\tau_2){\rm tr}\{\hat{n}({\bf r_2})G(t_f,\tau)\hat{A}G(\tau,t_f)\}
\label{eq:chi-RPA}
\end{eqnarray}
where $G(\tau_1,\tau_2)=G({\bf r_1},\tau_1;{\bf r_2},\tau_2)$ is the
one-particle contour Green's function, $W({\bf r_1},\tau_1;{\bf
  r_2},\tau_2)$ is a dynamically screened Coulomb interaction,
$\hat{n}({\bf r})$ is a one-particle density operator, and all traces
are taken over a one-particle Hilbert space.

Equation (\ref{eq:chi-RPA}) shows that for the practical calculation
of the correlation function $\chi^{C}_{\hat{A},\hat{V}}(t_f,\tau)$,
and thus the gradient of Eq.~(\ref{eq:gradient-contour}) we need the
countour ordered Green's function $G$ and the screened interaction
$W$. The latter is given by the RPA integral equation:
\begin{equation}
\setlength{\unitlength}{1pt}
\begin{picture}(220,30)
\put(0,0){
\includegraphics[clip]{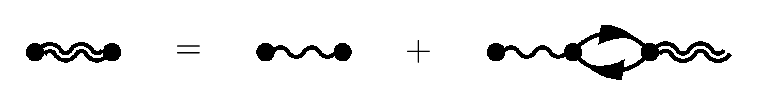}
}
\end{picture}
\end{equation}
while the former is calculated by propagating the Kadanoff-Baym equation:
\cite{KadBay},
\begin{equation}
 \label{eq:KBE}
\left(i\frac{\partial}{\partial \tau_1} - \hat{h}(1)\right)G(1,2) = \delta (1,2) +\int d3\Sigma(1,3)G(3,2),
\end{equation}
and its conjugate on the time contour. In Eq.~(\ref{eq:KBE})
$\hat{h}(1)=\hat{h}({\bf r_1},\tau_1)$ is the one-particle Hamiltonian
which also includes the Hartree potential, and the self energy is
given by the GW diagram:
\begin{equation}
\setlength{\unitlength}{1pt}
\begin{picture}(200,40)
\put(0,17){$\Sigma(1,2) = G(1,2)W(2,1) =$}
\put(120,0){
\includegraphics[clip]{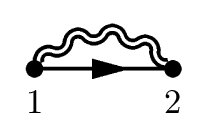}
}
\end{picture}
\end{equation}
More technical details can be found, for example, in
Ref.~\onlinecite{stan-2009}. At this point it is worth to comment on
one technical issue. Most currently existing implementations of the
Kadanoff-Baym equations \cite{DahStaLee2006,stan-2009,FriVerAlm2009}
assume that the dynamics starts from the thermal equilibrium state at
some temperature $T=1/\beta$. The equilibrium initial conditions are
technically convenient because they can be treated by a slight
modification of the Keldysh contour. Namely, one attaches a ``vertical
track'' going from $t_0$ to $t_0-i\beta$ from the backward branch of
the contour, and imposes antiperiodic Martin-Schwinger boundary
conditions $G(t_0-i\beta,\tau)=-G(t_0,\tau)$ on the Green's
function. If this formalism is employed then all time integrations in
Eqs.~(\ref{eq:chi-contour}) and (\ref{eq:KBE}) are along the modified
contour including the vertical track. However, this does not influence
the calculation of the gradient of Eq.~(\ref{eq:gradient-contour}) as
it requires only the correlation function on the real time, forwards
and backwards branches of the contour. We would like to emphasize that
the use of equilibrium/ground state initial conditions is not a
fundamental restriction of NEGFT. It is also possible to formulate the
theory for a general initial state
\cite{Hall1975,Danielewicz1984,Danielewicz1990,LeeSte2011arxiv}
although we are not aware of any practical implementation of this
formalism.

We conclude this section by noting the following remarkable fact
regarding the Keldysh contour formulation of QOCT for interacting
many-body systems. If the quantum dynamics is described within NEGFT
the implementation of QOCT does not require solving any additional
equation. All ingredients required to calculate the merit function
gradients are already known from the solution of the Kadanoff-Baym
equations. For example at the RPA/GW level of the theory one only
needs to plug the known functions $G$ and $W$ into
Eqs.~(\ref{eq:chi-contour}) and (\ref{eq:gradient-contour}), perform
the integrations, and close the optimization loop.

\section{Conclusions}

We have shown how the key equations of QOCT can be easily derived by
employing the formalism of linear response theory. These equations
provide the gradient of the target functional with respect to the
external field which has to be optimally shaped. In the light of the
linear response interpretation, the gradient is in fact the response
function of the driven system. First of all, this derivation is
valuable methodologically as it explains the internal structure of the
coherent control theory using one of the most common techniques in
theoretical physics, thus making QOCT more clear and accessible to a
broad audience. In addition to that our LRT representation immediately
suggests a reformulation of QOCT equations in terms of the Keldysh
contour-ordered correlation functions. The theory of non-equilibrium
Green's functions (NEGFT) may then be directly applied to derive new
approximation strategies for control problem in interacting
many-electron systems. We stress out that the implementation of QOCT
looks especially simple, if the quantum dynamics is described within
NEGFT, as it is frequently done in practice for many-body systems. To
calculate the merit function gradients there is no need to solve any
additional equation, since all the required quantities are already
known from the solution of the Kadanoff-Baym equations.  Work along
this line is in progress.

\begin{acknowledgments}
This work of IVT was supported by the Spanish MICINN (FIS2010-21282-C02-01), ``Grupos Consolidados UPV/EHU del Gobierno Vasco'' (IT-319-07), and the European Union through e-I3 ETSF project (Contract No. 211956).
\end{acknowledgments}

\appendix

\section{Generalized Kubo formula}
\label{appendix-lrt}

Let us consider a system governed by a total Hamiltonian $\hat{H}(t)$ that is split as:
\begin{equation}
\label{H}
\hat{H}(t) = \hat{H}_0(t) + f(t)\hat{V}\,,
\end{equation}
given some real time-dependent function $f(t)$ supported in the time
interval $[t_0,t_f]$.  To formulate a generalized LRT we need to solve
the equation of motion for the density matrix $\hat{\rho}(t)$
\begin{equation}
\label{EOM1}
 i\frac{\partial}{\partial t}\hat{\rho}(t) =[\hat{H}_0(t)+ f(t)\hat{V},\hat{\rho}(t)]
\end{equation}
for some given initial $\hat{\rho}(t_0)$, by considering the second
term as a ``perturbation'', while allowing the first term,
$\hat{H}_0$, to be time dependent.

We search for a solution in the following form
\begin{equation}
 \label{rho}
\hat{\rho}(t) = \hat{\rho}_0(t) + \hat{\rho}_1(t),
\end{equation}
where $\hat{\rho}_0(t)$ solves Eq.~(\ref{EOM1}) with $f(t)=0$ and the initial condition $\hat{\rho}_0(t_0)=\hat{\rho}(t_0)$, and $\hat{\rho}_1(t)$ is a solution to the linearized equation 
\begin{equation}
\label{EOM-lin}
 i\frac{\partial}{\partial t}\hat{\rho}_1(t) =[\hat{H}_0(t),\hat{\rho}_1(t)]+ [f(t)\hat{V},\hat{\rho}_0(t)]
\end{equation}
with the initial condition $\hat{\rho}_1(t_0)=0$.

It is convenient to introduce a propagator $\hat{U}(t,t')$ for the unperturbed evolution
\begin{equation}
\label{U}
\hat{U}(t,t') = \hat{T}e^{-i\int_{t'}^{t}d\tau\hat{H}_0(\tau)}\,,
\end{equation}
where $\hat{T}$ is the usual time-ordering operator. Equation~(\ref{U}) is a formal solution to the equations
\begin{eqnarray}
\nonumber
i\frac{\partial}{\partial t}\hat{U}(t,t') & = & \hat{H}_0(t)\hat{U}(t,t'), \,\,
\\\label{U-equation}
i\frac{\partial}{\partial t'}\hat{U}(t,t') & = & -\hat{U}(t,t')\hat{H}_0(t')
\end{eqnarray}
with the boundary condition $\hat{U}(t,t) = \hat{I}$.

Using Eqs.~(\ref{U-equation}) we immediately find both the unperturbed density density matrix $\hat{\rho}_0(t)$ and the solution $\hat{\rho}_1(t)$ of the linearized equation (\ref{EOM-lin}):
\begin{eqnarray}
 \label{rho_0}
&&\hat{\rho}_0(t) = \hat{U}(t,t_0)\hat{\rho}(t_0)\hat{U}(t_0,t),\\
 \label{rho_1}
\hat{\rho}_1(t) &=& -i\int_{t_0}^t d\tau \hat{U}(t,\tau)[f(\tau)\hat{V},\hat{\rho}_0(\tau)]\hat{U}(\tau,t).
\end{eqnarray}
It is easy to check that $\hat{\rho}_1(t)$ of Eq.~(\ref{rho_1}) is the solution to Eq.~(\ref{EOM-lin}). Indeed, the differentiation with respect to the upper limit of the $\tau$-integral in Eq.~(\ref{rho_1}) yields the second term in the right hand side in Eq.~(\ref{EOM-lin}), while the $t$-derivatives of the propagators in Eq.~(\ref{rho_1}) produce the first term, $[\hat{H}_0(t),\hat{\rho}_1(t)]$.

Now one can calculate the change $\delta A(t)$ of the expectation value for any observable $\hat{A}$, which is induced by the perturbation [the second term in the Hamiltonian (\ref{H})]:
\begin{equation}
 \label{deltaA-def}
\delta A(t) = {\rm Tr}\{\hat{\rho}_1(t)\hat{A}\}.
\end{equation}
Inserting $\hat{\rho}_1(t)$ of Eq.~(\ref{rho_1}) into Eq.~(\ref{deltaA-def}) and rearranging terms we get the result
\begin{equation}
\label{eq:kuboformula}
\delta A(t) =  -i\int_{t_0}^{t}d\tau\ f(\tau)
{\rm Tr}\left\{ \hat{\rho}(t_0)\left[ \hat{A}_H(t),\hat{V}_H(\tau)\right]\right\},
\end{equation}
where operators $\hat{O}(t)$ in the Heisenberg representation are defined as follows
\begin{equation}
 \label{O-Heisenberg}
\hat{O}_H(t) :=  \hat{U}(t_0,t)\hat{O}\hat{U}(t,t_0)\equiv\hat{U}^{\dagger}(t,t_0)\hat{O}\hat{U}(t,t_0).
\end{equation}
Equation (\ref{eq:kuboformula}) suggests the definition of the $(\hat{A},\hat{V})$ response function as:
\begin{equation}
\label{eq:lrdef}
\chi_{\hat{A},\hat{V}}(t,t') = -i\theta(t-t')
{\rm Tr}\left\{ \hat{\rho}(t_0)\left[ \hat{A}_H(t),\hat{V}_H(t')\right]\right\}
\end{equation}
so that:
\begin{equation}
\label{eq:deltaa}
\delta A(t) = 
\int_{t_0}^{\infty}d\tau\; f(\tau)\chi_{\hat{A},\hat{V}}(t,\tau)\,.
\end{equation}
The response function of Eq.~(\ref{eq:deltaa}) has the standard form
of Kubo's formula~\cite{kubo-1966}. The only minor difference is that
for a time-dependent unperturbed Hamiltonian $\hat{H}_0(t)$ the
Heisenberg operators, Eq.~(\ref{O-Heisenberg}), are defined via the
time-ordered exponential of Eq.~(\ref{U}).

Finally, we note that the QOCT equations can also be derived in
yet another different but equivalent manner by making use of the
following identity for the quantum mechanical propagator associated to
a Hamiltonian that depends on a parameter $\lambda$:
\begin{equation}
\frac{\partial }{\partial \lambda}\hat{U}_\lambda(t_f,t_0) = 
-i\int_{t_0}^{t_f}\!\!\!\!\!{\rm d}t\; \hat{U}^\dagger_\lambda(t,t_f)
\frac{\partial \hat{H}_\lambda}{\partial \lambda}(t)\hat{U}_\lambda(t,t_0)\,.
\end{equation}
With this identity, it is straightforward to compute the derivative of:
\begin{eqnarray}
\nonumber
G[u] & = & \langle \Psi[u](t_f)\vert\hat{A}\vert\Psi[u](t_f)\rangle
\\
&  = &  
\langle \Psi_0 \vert \hat{U}_u(t_0, t_f)\vert\hat{A}\vert\hat{U}_u(t_f,t_0)\vert \Psi_0\rangle \,,
\end{eqnarray}
where $\hat{U}_u(t,t_0)$ is the propagator determined by the Hamiltonian $\hat{H}[u](t)$.

\bibliography{qoct}

\end{document}